\documentclass[12pt]{article}

\usepackage{epsfig}

\begin{document}

\title{The rf control and detection system for {\em PACO}
 the parametric converter detector}

\author{Ph. Bernard$^1$ , G. Gemme$^2$\footnote{e-mail: gianluca.gemme@ge.infn.it} , 
R. Parodi$^2$ and E. Picasso$^3$ \\
{\it\small ${}^{1)}$CERN, CH-1211, Geneva 23, Switzerland} \\
{\it\small ${}^{2)}$INFN-Sezione di Genova, via Dodecaneso 33, I-16146 Genova, Italy} \\
{\it\small ${}^{3)}$Scuola Normale Superiore, Piazza dei Cavalieri 7, I-56126, Pisa, Italy}}

\date{}

\maketitle

\begin{abstract}
In this technical note the rf control and detection system for a detector of small harmonic displacements based on two coupled microwave cavities ({\em PACO}) is presented. The basic idea underlying this detector is the principle of parametric power conversion between two resonant modes of the system, stimulated by the (small) harmonic modulation of one system parameter. In this experiment we change the cavity length applying an harmonic voltage to a piezo-electric crystal. The system can achieve a great sensitivity to small harmonic displacements and can be an interesting candidate for the detection of small, mechanically coupled, interactions (e.g. high frequency gravitational waves).
\end{abstract}

\section{Introduction}

In this technical note we describe the rf control and detection system for a detector of small harmonic displacements based on two coupled microwave cavities ({\em PACO}). This experimental configuration was initially proposed by Bernard, Pegoraro, Picasso and Radicati \cite{bppr1}, \cite{bppr2} and later put in practice by Reece, Reiner and Melissinos \cite{rrm1}, \cite{rrm2}, and has been discussed in some detail in previous papers \cite{bgpp1}, \cite{bgpp2}, \cite{bgpp3}, \cite{bgpp4}. Here we just remind the basic principles underlying the detector operation.

The detector consists of an electromagnetic resonator with two resonant frequencies $\omega_s$ and $\omega_a$ both much higher than the characteristic frequency of the harmonic perturbation $\Omega$, and which satisfy the resonance condition $ | \omega_s - \omega_a | = \Omega$. In the scheme proposed in \cite{bppr1}, \cite{bppr2} the two resonant modes are obtained coupling two identical resonators; $\omega_s$ is the frequency of the symmetric (even) resonant mode, while $\omega_a$ is the frequency of the antisymmetric (odd) one. If some energy is initially stored in the symmetric mode, an harmonic perturbation can induce the transition of some energy to the initially empty level that can be extracted at frequency $\omega_a$. The electromagnetic power in the antisymmetric mode is proportional to the square of the amplitude of the perturbation. 

In this experiment we use two cylindrical cavities coupled trough an axial iris and we change the cavity length applying an harmonic voltage to a piezo-electric crystal. To increase the sensitivity of the detector a resonator geometry and field configuration with high geometrical factor and high quality factor $Q$ are preferred. To avoid electron field emission from the cavity surface rf modes with vanishing electric field at the surface are mandatory. For these reasons we have chosen TE mode (TE$_{011}$) superconducting cavities. The choice of frequency was imposed by the maximum dimension of the resonator that can be housed in a standard vertical cyostat; in our case the inner diameter is 300 mm, giving us enough room for a 3 GHz, TE$_{011}$ resonator. 

The system can achieve a great sensitivity to small harmonic displacements (up to $\delta l / l \approx 10^{-20}$ @ $\Omega=$1 MHz, and can be an interesting candidate for the detection of small, mechanically coupled, interactions (e.g. high frequency gravitational waves).

\section{{\em PACO} rf system}

\begin{figure}[t]
\begin{center} \mbox{\epsfig{file=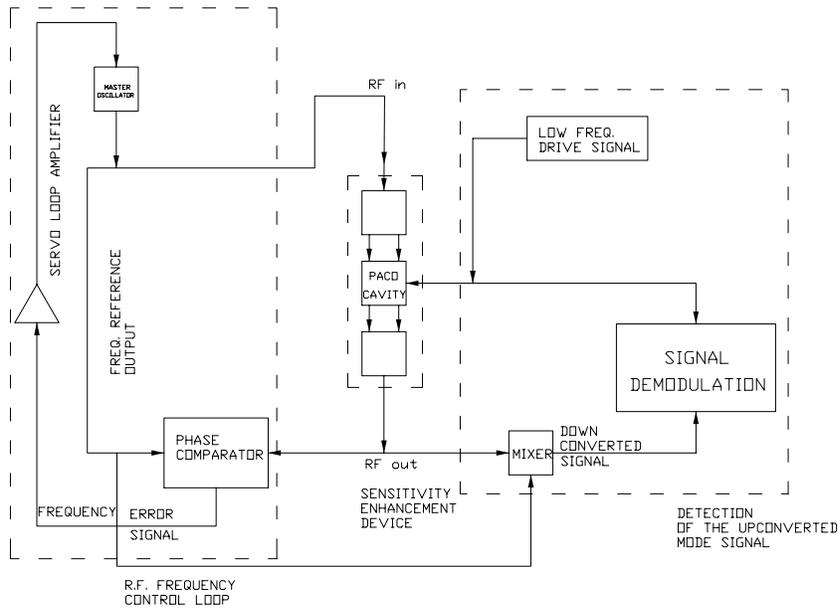}} \end{center}
\caption{{\em PACO} rf system} 
\label{general} 
\end{figure}

The three main functions of rf control and measurement system of the {\em PACO} 
experiment are shown in figure \ref{general}.

The system performs three separate functions listed in the following.
\begin{enumerate}
\item{The first task of the system is to lock the rf frequency of the master 
oscillator to the resonant frequency of the even mode of the cavity and to keep 
constant the energy stored in the mode.}
\item{The second task is to increase the detector's sensitivity by driving the 
coupled resonators purely in the even mode and receiving only the rf power up-
converted to the odd mode by the perturbation of the cavity walls. This goal can be 
obtained by rejecting the signal at the even mode frequency tacking advantage of the 
symmetries in the field distribution of the two modes.}
\item{The third task is the detection of the up-converted signal pushing the 
detector's sensitivity to the limit set by the contribution of the noise sources at the 
operating frequency. The various noise sources have been described and discussed in a previous paper \cite{bgpp2}.}
\end{enumerate}
 
\section{The rf control loop}

The output of the master oscillator  (HP4422B) is fed into the cavity trough a TWT 
amplifier giving a saturated output of 20 Watt in the frequency range of  2-4 GHz.
The stored energy in the cavity is adjusted at the operating level by controlling the 
output of the master oscillator via the built in variable attenuator. 

The output signal is sampled via a 3 dB power splitter. The $A$ output of the 
splitter is sent to the TWT amplifier, the $B$ output is sent, trough the phase shifter (PS), to 
the local oscillator (LO) input of a rf mixer acting as a phase detector (PD).
The output of the rf power amplifier is fed to the resonant cavity trough a 
double directional coupler, and a $180^{\rm o}$ hybrid ring acting as a magic tee.
The rf power enters the magic tee via the sum arm, $\Sigma$, and is split in two signals 
of same amplitude and zero relative phase, coming out the tee co-linear arms 1 and 2.
 
The rf signal, reflected by the input ports of the cavity, enters the magic tee 
trough the co-linear arms. The two signals are added at the $\Sigma$ arm and sampled by the 
directional coupler to give information about the energy stored in cavity allowing for 
the measurement of the coupling factor, quality factor, stored energy. 
While driving the cavity on the even mode no reflected signal is shown at the $\Delta$ port of 
the magic tee where the signals coming from the co-linear arms are algebraically added 
to zero due to the $180^{\rm o}$ phase shift between the two signals coming by the co-linear 
arms.

To get the maximum of the performances of the magic tee we need to have equal 
reflected signals (phase and amplitude) at the cavity input ports.
The equal amplitude goal can be achieved by a careful design of the 
input couplers; a good design guarantees us also very nice phase equality at the ports.
To preserve the signal integrity we use matched output lines (in phase and 
amplitude) inside the cryostat.
The cable used for the output lines is the best rf cable (as far as attenuation and 
phase stability are concerned) money can buy, nevertheless no characterisation at 
cryogenic temperatures exists.
Because the phase shift is very sensitive to temperature inhomogeneities between 
the two cables and the phase difference between the two co-linear arms of the magic tee 
gives a quite strong signal at the $\Delta$ port, we need to compensate for 
differential thermal contractions of the cables inside the cryostat, leading to phase 
unbalance in the feed lines.
To do that we insert a phase shifter in one of the lines to reduce to a minimum the 
leakage of the unwanted modes on the two ports.
As we will show in the following section, mode leakage of the even mode to the 
$\Delta$ port sets a limit to the system sensitivity increasing the overall noise level of the 
detector.

Mode leakage of the odd mode to the $\Sigma$ port reduces the system sensitivity by 
reducing the signal level available for detection.
The converted rf power (odd mode) coupled to the input $\Sigma$ port is lost forever.
The rf system is symmetric on both the input and output ports of the detector 
cavity. 
The output ports of the cavity are coupled for a maximum output signal 
on the odd mode (detection mode) and the magic tee is used to reject the rf power at 
the frequency of the even mode.
The up-converted signal (odd mode) comes out at the $\Delta$ port of the magic tee.
A fraction of the signal at the $\Sigma$ port is fed to the rf input of the phase detector 
PD via a low noise rf amplifier.
The intermediate frequency  (IF) output of the phase detector PD is fed back to 
the rf master oscillator to lock the output signal to the resonant frequency of the 
resonator.
The total phase shift around the loop is set trough the phase shifter PS, to have the 
maximum of energy stored in the detector.
A carefully design of the servo loop amplifier (SLA) guarantees the stability of 
the system and the rejection of the residual noise of the master oscillator up to one 
MHz.
The same fraction of the $\Sigma$ output of the output magic tee is used to keep 
constant, to 100 ppm, the energy stored in the cavity feeding back an error 
signal to drive the electronically controlled output attenuator of the master oscillator.

Great deal of care is needed in tailoring the frequency response of both controls 
because the two loops can interact producing phase-amplitude oscillations in the rf 
fields stored in the cavities.

\begin{figure}[t]
\begin{center} \mbox{\epsfig{file=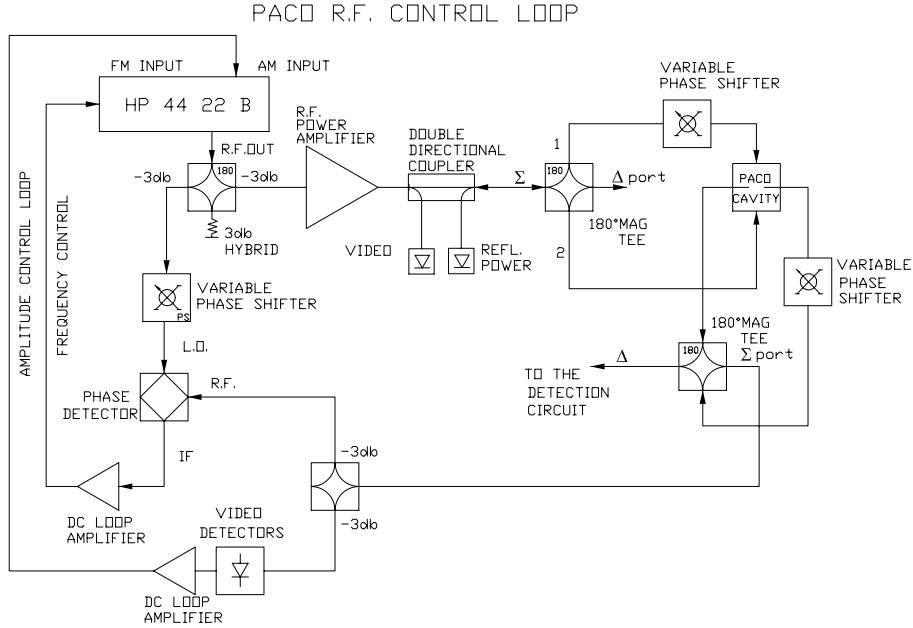}} \end{center}
\caption{{\em PACO} rf control loop} 
\label{loop} 
\end{figure} 

\section{Sensitivity enhancement using the mode symmetry}
 
The two modes of the detector cavity have (as in the case of two coupled 
pendulums) opposite symmetries of the fields. 
The rf detection system can greatly be improved if the mode symmetry 
information is used in the rf feed and detection system to reduce the residual 
components of the noise produced by the master oscillator line width.
A further improvement is obtained tacking advantage of the very high selectivity 
of the resonator's modes having a quality factor $Q  \approx 10^{10}$.

\begin{figure}[t]
\begin{center} \mbox{\epsfig{file=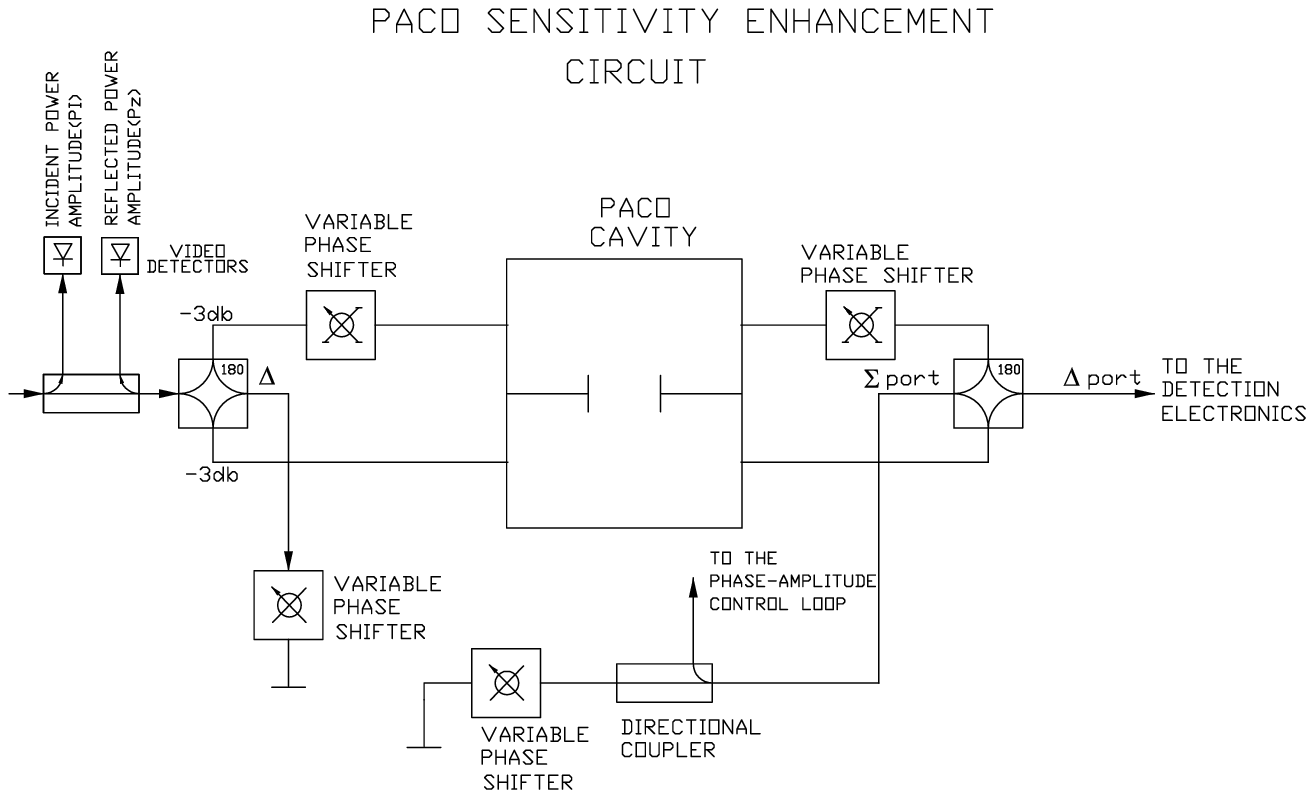}} \end{center}
\caption{Sensitivity enhancement circuit} 
\label{sensit} 
\end{figure} 

Using two separate sets of ports to drive the cavity and to receive the up-converted
signal (at the frequency of the odd mode), the cavity acts as a very sharp 
filter, with an high rejection of the signal noise (coming from the master oscillator) at 
the frequency of the up-converted signal. 
The resulting attenuation is given by the shape of the cavity resonance, a Lorentz 
shape with a full width half maximum $FWHM = f/Q$, where $f$ is the resonant frequency.
This already low residual noise, can be even more reduced with the two magic tees.
Using  two equal couplers at the cavity input, driven by the two co-linear arms of a 
magic tee fed via the $\Sigma$ port, we store energy in the cavity only on the even mode.
Receiving the up-converted signal at  the $\Delta$ port the of  the second magic tee, rejects 
the even mode components from  the cavity by an amount given  by the magic tee 
insulation.

In the case of an ideal magic tee the mode rejection is infinite as the tee insulation. 
No even mode component is transmitted trough the system and there will be no signal
at the output port if the cavity is driven purely in the even mode.
In the ideal case this results is obtained also in the more simple scheme used 
by Melissinos and Reece \cite{rrm1}, \cite{rrm2}, measuring the up converted power coming out of the detector 
along the input lines.

Our scheme gives better sensitivity and performances in the real case.
The first obvious gain is the sum of the $\Delta$ to $\Sigma$ port insulation of the two tees, plus 
the possibility of adjusting separately the input and output lines to get better mode 
rejection. In a commercial magic tee the insulation is specified to be 
$\approx 25 - 35$ dB over the whole bandwidth of the device (usually an octave in 
frequency).
The reason for this quite low insulation is mainly due to the difficulty of balancing 
on a large range of frequency the phases of the signals coming from the two co-linear 
arms of the tee.

Consider two equal signals in phase 
\begin{equation}
V = V_0 \cos( \omega t + \phi )
\end{equation}
entering the co-linear arms of  the magic tee; suppose a phase unbalance of  $\alpha$ 
degrees between the two path to the $\Delta$ port : the resulting signals at the $\Delta$ are
\begin{eqnarray}
V_{\Delta} = V_0 \cos ( \omega t + \phi ) - V_0 \cos ( \omega t + \phi + \alpha )  = \nonumber \\
= V_0 \cos (\omega t + \phi ) - V_0 \cos (\omega t + \phi )  \cos \alpha + V_0 \sin (\omega t + \phi ) \sin \alpha
\end{eqnarray}
or, for a small value of $\alpha$
\begin{equation}
V_{\Delta}=\alpha \, V_0 \sin (\omega t + \phi ) 
\end{equation}

Now a phase unbalance as small as five degrees reduces the insulation from $\Delta$ to $\Sigma$ 
port to only 25 dB.
This fairly low insulation in magic tees comes from the difficulty of designing a 
90 degrees rf broad band phase shifter.
A custom designed magic tee can be optimized to give a better insulation (35-40 
dB) on a more limited frequency interval. 
Since the bandwidth-insulation product is roughly constant, 
pushing the requirements to high insulation (70-80 dB) reduces too much the useful 
bandwidth, giving not so much flexibility in the rf system design .
Any mismatch between the frequencies of the cavity modes and of the magic tees 
results in a very severe degradation of the noise rejection for the whole system at the 
detection frequency, spoiling the ultimate sensitivity of the detector.

Our electronic scheme allows for an independent compensation of the magic tee 
phase mismatch either at the feed frequency and at the detection frequency in a flexible 
way: the phase mismatch is compensated using a variable phase shifter at the input 
of one of the co-linear arms.
Getting the optimum phase at the input side will results in a pure excitation of the 
even (drive) mode of the two cavity system, keeping the power at the frequency of the 
odd (detection) mode 70 dB (the tee insulation) below the level of the drive mode.
Adjusting the phase at the output will couple to the output only the odd mode 
components rejecting the even mode component by 70 dB.
The total even-odd mode rejection of the system is the sum of the attenuation we 
can obtain from the two $90^{\rm o}$ hybrids; in the real world a practical limit is set by the cross 
talk effects between input and output limiting to 120-130 dB the maximum achievable 
insulation.

The price we pay for the mode selectivity improvement in our set-up is affordable 
and gives us the possibility of a separate and more controlled tuning of the input and 
output circuits.
The use of two separated ports for the input and output (with separated magic 
tees) add some complication to the rf system.
The coupling coefficient of the input and output port of the two cell cavity need to 
be critically coupled ($\beta$ = 1) to the rf source and the rf detection system.
In this way we will have the optimum transfer of power to the even mode (a 
maximum of stored energy) and to the odd mode (a maximum in the detector output).
Because the frequency and field distribution of the two modes are quite close, the 
input and output ports will be critically coupled to both modes. For that reason 50$\%$
of the even mode signal will be coupled to the idle $\Sigma$ port at the output magic tee, 
and symmetrically 50$\%$
of the odd mode signal is coupled to the idle $\Delta$ port at the input magic tee.
The new couplings will greatly affect the detector's performances, affecting the 
loaded quality factor of the cavity and changing in a substantial way the way to couple 
the cavity to the rf source.

At the glance it is clear that closing the two idle ports with a matched load will 
worsen the detector performance due to the following reasons:
\begin{enumerate}
\item{Half of the converted power at the odd mode frequency will be lost 
worsening of a factor 2 the detector sensitivity.}
\item{The loaded quality factor of the cavity is reduced by a factor roughly 2 with 
the same reduction in the system sensitivity.}
\item{The coupling to the rf generator is reduced with a reduction of the detector 
efficiency (due to the input mismatch the incoming power is partially reflected 
and more power is needed from the rf system to store the same energy in the 
cavity).}
\end{enumerate}

A detailed analysis showed that closing the idle ports with an impedance giving 
a reflection coefficient amplitude one and phase zero solves the problem reflecting back 
to the cavity all the rf power coupled to the idle ports.
Using that termination of the idle ports the two rf arms at the input and output 
are de-coupled (at least by the mode selectivity factor equal to the sum of the $\Delta$ and $\Sigma$ ports 
insulation of the two hybrids).
An rf impedance having reflection coefficient equal to one is an open circuit or a short 
circuit plus a $180^{\rm o}$ phase shift.
For this reason we close the two idle ports using a phase shifter and a short.
Again this solution gives the possibility to fine tune the system and compensate 
for the effect of non-ideal elements in the rf system.

At the $\Sigma$ port of the detection arm we insert a directional coupler to sample 
a tiny amount of the even mode power coming from the cavity.
This transmitted power is fed to the frequency-amplitude servo loop used to lock 
the frequency of the master oscillator to the cavity frequency and to keep constant 
the energy content in the cavity.
This choice gives to our rf control system a better reliability over the 
Melissinos-Reece control system using the reflected power from the cavity for the 
frequency control loop.
The effect of this sampling behaves in our system as the sampling antenna used 
to monitor and control the rf power fed to a superconducting cavity.   

As a final remark our system, despite some complexity, guarantees the following 
improvement over the one used in the previous experiment:
\begin{itemize}
\item{A better rejection of the phase amplitude noise of the master oscillator obtained 
using the sharp resonance of the resonator itself.}
\item{A better insulation of the drive and detector ports obtained by using separate 
drive and detection arms of the rf system.}
\item{The possibility of an independent adjustment of the phase lag in the two arms 
giving a better magic tee insulation at the operating frequencies.}
\item{A greater reliability for the frequency amplitude loop using the transmitted power, 
instead than the reflected, coming from the cavity.}
\end{itemize}

\section{Detection of the converted signal}

\begin{figure}[t]
\begin{center} \mbox{\epsfig{file=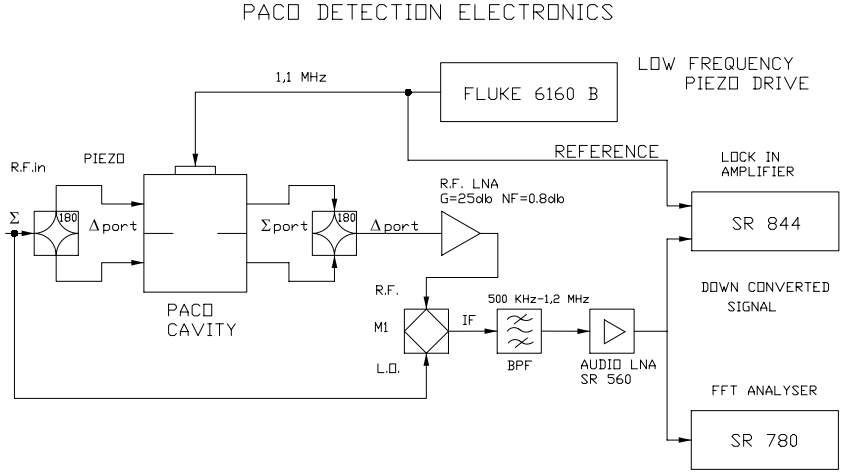}} \end{center}
\caption{{\em PACO} detection system} 
\label{detection} 
\end{figure} 

The signal converted to the odd mode by the interaction between the mechanical 
perturbation and the rf fields is coupled to the $\Delta$ port of the detection arm of the rf system 
and amplified by the low noise rf amplifier LNA.
Great deal of care must be used in the choice of the LNA, because the noise figure of 
the amplifier greatly affects the detector's sensitivity in the detection region above the 
mechanical resonances of the cavity (typically spanning the frequency interval 1-10 kHz)
The LNA we choose is a commercial, room temperature, low noise amplifier with a 48
dB gain and quite large bandwidth of 500 MHz centered on our operating frequency of 
3 GHz.
The LNA Noise figure is 0.8 dB corresponding to a noise temperature of 360 K.

The converted rf signal amplified by the LNA is fed to the rf input port (RF) of a 
low noise double balanced mixer M1; the local oscillator port (LO) of M1 is driven by the 
even mode rf power (at frequency $\omega_s$) transmitted by the cavity. The LO input level is 
adjusted for a minimum noise contribution from the mixer.
As shown in the previous section the input spectrum to the RF port of the mixer is 
composed by two signals:
the first at frequency $\omega_s$ coming from the rf leakage of the even mode trough the 
detection system (even if greatly reduced by the aforementioned double rejection of the 
carefully tuned magic tees); the second is the converted energy on the odd mode at frequency 
$\omega_a$.
Both signal are down converted by the M1 mixer giving to the IF port a DC signal 
proportional to the even mode leakage and the signal at frequency $\Omega$ proportional to the odd 
mode excitation.

The down-converted IF output is further amplified using a low noise audio 
preamplifier (Stanford Research SR560).
The combination of tunable built in filters of the SR560 amplifier allows us to further 
reject (if necessary) the DC component coming from the even mode leakage.
Last the output of the audio amplifier is fed to the lock in amplifier (Stanford 
Research SR844) used as synchronous detector driven by the low frequency synthesizer used 
to drive the detector cavity trough a piezoelectric ceramic at frequency $\Omega$.

The detection electronics for the detection of harmonic, mechanically coupled interactions, at frequency $\Omega$ (as 
gravitational waves) is slightly different.
Since the exact frequency and phase of the driving source is not known, we 
can't perform a synchronous detection; we need to perform a auto 
correlation of the detector output - or to cross-correlate the outputs of two different detectors - to detect the down converted component at $\Omega$.
The output of the SR560 preamplifier is fed to a fft signal 
analyzer (Stanford Research SR780) able to average the input signal on a bandwidth as small 
as 0.1 mHz, giving us a very comfortable margin to get the ultimate sensitivity foreseen for 
the feasibility study of our detector.

The outlined scheme of electronic detection gives us the non-marginal benefit of being 
(at the first order at least) self compensating against perturbations changing in the same way 
the frequency of the two modes of the detector.
This type of perturbation is usually produced by changes of the cavity walls due to 
changes of pressure of the helium bath used to keep the cavity at the operating 
temperature or by the thermal contraction of the cavity walls produced by changes of the 
operating temperature.
Furthermore similar effects are produced by the radiation pressure of the 
electromagnetic energy stored in the even mode; this pressure acts in a symmetrical way on 
the cavity walls producing equal deformation of both the detector's cells.
The result of the deformation is a frequency shift of both the modes of a same amount 
of frequency; this kind of deformation does not affect (at the first order at least) the coupling of the 
cells fixing the amount of splitting between the even and odd modes.
As a result the frequencies of the two modes $\omega_s$ and $\omega_a$ will change under the effect of 
pressure changes on the cavity walls but the difference $\Omega$ (related to the coupling) will not.

In our scheme the local oscillator (used for the down conversion of the rf signal 
coming from the detector) is the rf signal transmitted by our cavity at the frequency of the 
even mode.
In this way the common mode frequency drift of the resonators is automatically 
cancelled (at the first order at least) and the mixer M1 output is a signal exactly at the 
perturbation frequency $\Omega$.
This effect gives us the possibility of using a quite simple refrigeration scheme (without 
any complex and cumbersome servo loop) to keep constant the helium bath temperature and pressure.

\section{Experimental results}

The electromagnetic properties of the cavity have been measured in a vertical cryostat after careful tuning of the two cells frequencies.
The symmetric mode frequency was $\omega_s/(2\pi) = 3.03431\,$ GHz and the mode separation was $\Omega = 1.3817\,$ MHz.
The unloaded quality factor at 4.2 K was $Q = 5 \cdot 10^7$, and no significant improvement was found lowering the helium bath temperature at 1.8 K. Even after a second chemical polishing, performed at CERN, which removed approximately $300 \,\rm{\mu m}$ of niobium from the surface, no improvement was observed. We believe that this very low $Q$ value is due to hot spots on the surface caused by welding problems occurred during the cavity fabrication.  Adjusting the phase and amplitude of the rf signal entering and leaving the cavity, the arms of the two magic tees were balanced to launch the even mode at the cavity input and to pick up the odd mode at the cavity output. With 30 dBm of power at the $\Sigma$ port of the first magic tee, -90 dBm were detected at the $\Delta$  port of the second one, giving an overall attenuation of the symmetric mode of 120 dB. 
The energy stored in the cavity with 30 dBm input power was approximately 1 mJ.
The signal emerging from the $\Delta$ port of the output magic tee was amplified by the LNA and fed into a spectrum analyzer. The signal level at frequency $\omega_a$ was –120 dBm in a 1 Hz bandwidth.
System sensitivity at this stage is given by
\begin{equation}
h_{min} \approx 6.55 \cdot 10^{-18} \, {\rm Hz^{-1/2}}
\end{equation}

This value is quite far from our goal of $h_{min} = 10^{-20}$ (Hz)$^{-1/2}$. Since, to our knowledge, this is the first example of a parametric detector operated in trasmission, and since this configuration requires very careful adjustments of the input and output ports balancing, we believe that significant improvements are obtainable. Furthermore the new cavity under construction should show the high quality factor needed to reach high sensitivity.

\end{document}